\documentclass[10pt,journal,compsoc]{IEEEtran}
\usepackage{graphicx}
\usepackage{algorithm, algorithmic}
\usepackage{amsmath}

\usepackage{float}

\ifCLASSOPTIONcompsoc
  \usepackage[nocompress]{cite}
\else
  \usepackage{cite}
\fi
\ifCLASSINFOpdf
\else
\fi
\begin{document}
%
\title{Blockchain-enabled Authentication Handover with Efficient Privacy Protection in SDN-based 5G Networks}
%
%
%
%

\author{Abbas Yazdinejad,
        	Reza M. Parizi,~\IEEEmembership{Senior Member,~IEEE,}
        	Ali Dehghantanha,~\IEEEmembership{Senior Member,~IEEE,}
        and~Kim-Kwang Raymond Choo,~\IEEEmembership{Senior Member,~IEEE}
\IEEEcompsocitemizethanks{\IEEEcompsocthanksitem A. Yazdinejad, was with the Faculty of Computer Engineering, University of Isfahan, Iran. Email: abbasyazdinejad@yahoo.com.\protect\\
\IEEEcompsocthanksitem R.M. Parizi is with the Department of Software Engineering and Game
Development, Kennesaw State University, GA 30060, USA. Email:
rparizi1@kennesaw.edu.
\IEEEcompsocthanksitem A. Dehghantanha is with the Cyber Science Lab, School of Computer Science,
and University of Guelph, Ontario, Canada. Email: adehghan@uoguelph.ca
\IEEEcompsocthanksitem *K.K.R. Choo is with the Department of Information Systems and Cyber
Security, University of Texas at San Antonio Texas, USA. Email:
raymond.choo@fulbrightmail.org (Corresponding author)}
\thanks{}}

%
%

\markboth{IEEE TRANSACTIONS JOURNAL, 2019}%
{Shell \MakeLowercase{\textit{et al.}}: Bare Demo of IEEEtran.cls for Computer Society Journals}
%



\IEEEtitleabstractindextext{%
\begin{abstract}
		5G mobile networks provide additional benefits in terms of lower latency, higher data rates, and more coverage, in comparison to 4G networks, and they are also coming close to standardization. For example, 5G has a new level of data transfer and processing speed that assures users are not disconnected when they move from one cell to another; thus, supporting faster connection. However, it comes with its own technical challenges relating to resource management, authentication handover and user privacy protection. In 5G, the frequent displacement of the users among the cells as a result of repeated authentication handovers often lead to a delay, contradicting the 5G objectives. In this paper, we propose a new authentication approach that utilizes blockchain and software defined networking (SDN) techniques to remove the re-authentication in repeated handover among heterogeneous cells. The proposed approach is designed to assure the low delay, appropriate for the 5G network in which users can be replaced with the least delay among heterogeneous cells using their public and private keys provided by the devised blockchain component while protecting their privacy. In our comparison between Proof-of-Work (POW)-based and network-based models, the delay of our authentication handover was shown to be less than 1ms. Also, our approach demonstrated less signaling overhead and energy consumption compared to peer models.
\end{abstract}

}

\maketitle

\IEEEdisplaynontitleabstractindextext

%
\IEEEpeerreviewmaketitle

\IEEEraisesectionheading{\section{Introduction}\label{sec:introduction}}

%
%
%
%
\IEEEPARstart{T}{he} rapid growth of mobile devices and applications along with their processing needs have led to the emergence of the fifth generation (5G) networks. The 5G networks have been introduced with properties like higher bit rate than 10 Gb/s, low latency and increased network coverage compared to 4G \cite{m1}. The 5G networks operate through heterogeneous cells and expand overlay coverage \cite{m2,m3}. The 5G users, like Internet of Thing (IoT) devices, vehicles, and mobile nodes, when moving from one cell to another, make the handover process activated and if this handover is frequently run, it could lead to a delay in the 5G network \cite{m4}. Following a frequent handover, the authentication mechanisms become more involved and may increase the delay time, which contradicts the 5G objectives. Using inefficient authentication handover could cause performance degradation among heterogeneous 5G cells and increases the delay. The power and resource constraints among the Access Points (APs) in cells require low complexity and highly efficient handover authentication procedures among heterogeneous and homogeneous cells in 5G \cite{m5}. The 5G architecture offers advantages in communication but has associated technical challenges including, authentication handover, the existence of heterogeneous cells, and privacy protection \cite{m5,m6}. Providing network management and security services inside heterogeneous cells can be challenging since mobile users (MU) may leave one cell for another frequently, and specifically when they are dealing with financial and data-sensitive applications.

5G requires taking into account the acceptable level of security in application scenarios and network architecture, especially in validating the facilities and providing their access level to 4G. Also, reducing the delay is one of the objectives and characteristics of 5G. An approach to validate and protect privacy, which is faster, safer and more effective, is essential for the advancement of the 5G networks. For 5G, the security requirements are higher in comparison with previous networks (2G, 3G, and LTE) for which new solutions are required to provide intelligent control across heterogeneous cells for reliable mechanisms and further adoption of the 5G network \cite{m3,m5}. Recent advanced technologies like SDN/NFV and blockchain \cite{m7,m8} have recently received a lot of attention for the advancment of the next generation of wireless networks. 

In SDN, the control plane is separated from the data plane, and its controlling part can meet the controlling needs in 5G, as the SDN is a new architecture of the network, with properties like programmability and flexibility in network’s management for testing new ideas \cite{m9,m10}. The infusion of SDN in 5G is beneficial because, in the future, mobile networks would be going towards more scalability requiring better management and flexibility \cite{m9}. SDN flexibility could potentially benefit 5G applications, in terms of quality of service (QoS), machine to machine (M2M) and human to human (H2H) communications \cite{m9}. Applying blockchain, on the other hand, can also allow us to better respond to some of the security challenges in 5G \cite{m7}. Specifically, a blockchain is a fraud-resilience, distributed ledger that records all transactions in a 2P2 network. The blockchain has a decentralized architecture, and its popularity in cryptocurrency world in securing distributed communication has been remarkable \cite{m11}. Blockchain can play an important role in facilitating secure communication between mobile users in 5G, for example by removing intermediaries for authentication, reduction in transaction cost, and global accessibility for all users \cite{m12}. In other words, we posit that SDN and blockchain can be combined to facilitate us to provide enhanced privacy protection and security in 5G. Bblockchain technology is more widely known in financial \cite{m13} and supply chain applications\cite{m14}, but its adoption is limited on 5G mobile networks and mobile services because of its resource-intensive consensus and validation protocols, mainly Proof of Work (POW)  \cite{m15}. POW requires significant energy and processing time, which is not appropriate for mobile users, especially on the 5G network \cite{m15}. Having said that, the blockchain should be compatible with the 5G specification to be effective, and thus, we use an upgraded Delegated Proof of Stake (DPOS) algorithm in this work. 

In this paper, a blockchain-enabled authentication handover approach with effective privacy protection is proposed for the 5G within SDN platform. In the proposed approach, users obtain a quick and secure connection by eliminating re-authentication among handovers operators between heterogeneous cells with a low delay. Demonstrated in experimental results, the comparison between authentication delay with POW-based and network-based models, and the delay of our authentication handover shown to be less than 1 ms, making the proposed approach well suited for 5G. In addition, our approach showed to have less signaling overhead with better energy consumption in compared with POW-based and network-based models. This was achieved by applying the upgraded DPOS algorithm, which has been shown to be scalable and energy-optimized and effective in reducing latency as opposed to POW.

The rest of the paper is organized as follows: The proposed architecture of 5G with a heterogeneous cell of SDN and blockchain on the 5G is described in Section 2. The blockchain-enabled authentication handover mechanism for 5G is presented in Section 3. The effective privacy protection in SDN based 5G network is discussed in Section 4. Section 5 defines the DPOS algorithm used in blockchain component. The simulations results and evaluation are presented in Section 6.  Section 7 presents the related work and finally, Section 8 gives the concluding remarks.

\section{The Proposed Architecture of 5G with a heterogeneous Cell}

In the proposed architecture, the blockchain and SDN are introduced into the 5G network to simplify and eliminate the frequent handover authentication into small cells and heterogeneous networks. The overall model of architecture is shown in Fig. 1. 
\begin{figure*}
	\centering
	\includegraphics[scale=0.050]{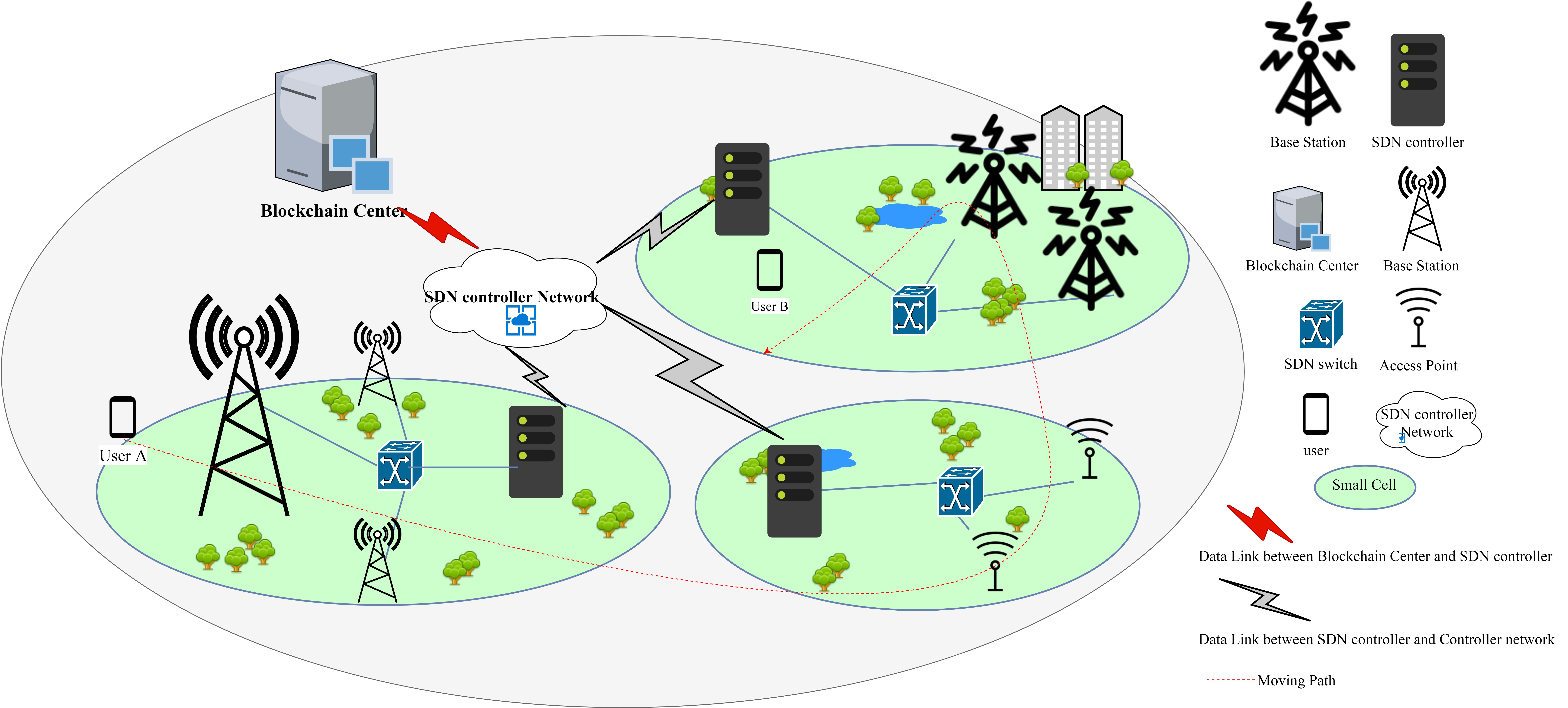}	
	{\scriptsize \caption{  The proposed architecture of blockchain-enabled handover in SDN-based 5G networks with heterogeneous cell }}
\end{figure*}
The blockchain center (BC) is located in an environment outside and near the cells or heterogonous cells and is applied as a space for storing and producing an encryption of parameters like the device identification, certification, and unique data related to the device under the alias. Encryption materials are applied to protect privacy and security. BC is involved in this architecture as follows: 1) the initial registration, the new devices require initial registration when they first want to enter a cell or heterogeneous network, 2) changing identity information, devices may change their alias when passing through a cell to another, thus their encrypts, therefore, BC generates new encryption for identification and validation in other cells, and 3) Hostile cancelation, in BC, malicious behaviors are detected by using blockchain lookup. The identity of the adversary is publicized once the malicious behaviors have been confirmed by BC among Cells. The BC is a public ledger that any MU can register into it, which is approved and recognized by mobile operators. Only MU who are known to mobile operators have permission to register in BC. In our architecture, we used a public blockchain to facilitate collaboration among various mobile operators who can select and approve associated MU with the chain in BC. In the blockchain, any MU can be part	of the network and can have their own private and public keys.
	Moreover, all SDN controller candidates can be involved in consensus mechanism and can also check and validate all transactions within the 5G network.

Heterogeneous network management and flexibility of the SDN structure is applied in the proposed architecture to increase programmability. The SDN controller is prescribed for the overall control cell or heterogeneous network. The SDN switch deals with the data transfer and behavior change in the network by following the controller commands. By separating the data plane and the control plane in the SDN, defining the protocol, functions, and policy on the 5G network becomes possible [9,10]. The SDN controllers in this proposed approach are one network and can communicate with each other and BC, and like Bitcoin, the information inside secure messages is exchanged as encapsulated transactions among them. Transactions and messages from BC can be shared through the dedicated transfer keys to the controller. Each SDN controller has a dedicated transfer key received from BC and is applied to transfer and receive information.
Scalability is an important problem in SDN, and we have solved it through a hierarchical structure between SDN controller and BC in our architecture. Not only the SDN controllers are one network that can communicate with each other (as another layer) but also they are being managed via BC (as a higher layer). If any SDN controller becomes down in a cell, the system will then manage this cell via another SND controller in the network between SDN controllers. In this architecture,  it is assumed that the data center of the mobile operator has the information of BC and SDN controllers can be controlled by mobile operators. The objective is to achieve an effective, secure and fast mechanism for authentication handover among the 5G network. In our BC, we have applied the optimal DPOS algorithm  \cite{m16} for energy consumption and speed boosting during transactions.
\section{Blockchain-enabled Authentication Handover Mechanism for 5G}%
By applying BC and SDN, a handover mechanism is designed for transmitting the key for authentication in the 5G with eliminating re-authentication among handovers between cells. Fig. 2 shows the relationship between BC, SDN, and AP/BS in our approach, and gives the main componets forming BC including a ledger (to store and maintain data), Auth\_Control and Sec\_info units.

\begin{figure}  [ht]
	\centering
	\includegraphics[scale=0.068]{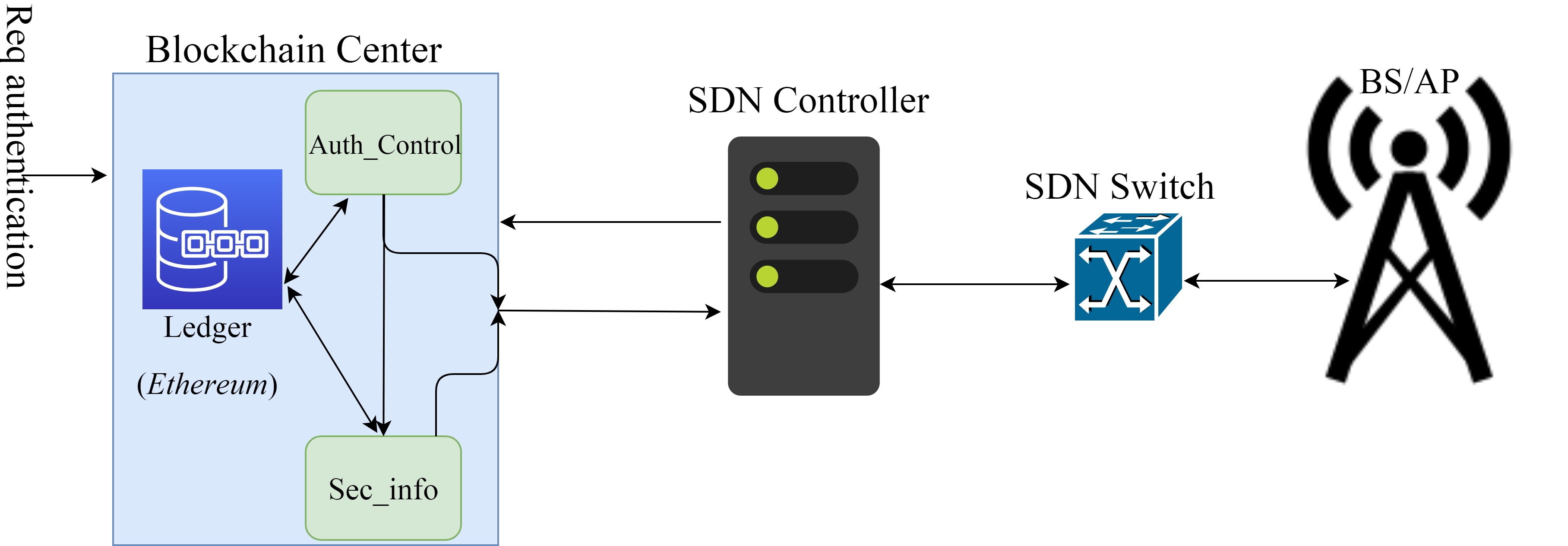}	
	{\scriptsize	\caption{ {\small Relationship among BC and SDN and AP/BS structure in the proposed 5G network} }}
\end{figure}
Initially, the mobile users (MU) are registered in BC to generate encryption material based on their properties and then receive the key and encryption properties, and BC sends a vector containing the MU information to the SDN of that cell in a simultaneous manner. BC assigns two public and private keys to each user’s mobile. As shown in Fig. 3, the mutual authentication between user and BC occurs in procedure 1. The MU sends the request of joining the BC to enter the given cell or domain. This request is received by the\textit{ Auth\_Control} unit in BC and sent to the MU after it is confirmed; meanwhile, another unit in BC named \textit{sec\_info} sends the data of this MU to the SDN controller in that domain or cell.

In this context, the \textit{Auth\_Control} inside BC is applied to identify the unique MU information like identity, location, direction, physical layer properties, RTT, and public and private key assignment.

Through the \textit{sec\_info} unit, the messages are sent to the SDN controller safely in capsuled packages. Specifically, a set of information from the registered user in BC like the public key is sent to the SDN controller of that and other cells. 

The SDN controller is responsible for MU management and defines the invoices for the SDN switch tables in a cell. It also stores the APs in the cell to validate the key, and the MU, and then the MU joins the cell. If for any reason the MU wants to handover from the existing AP to another AP in the same cell, the existing AP makes the SDN controller arena and the MU sends the associate request to the target AP and then becomes disconnected with the existing AP.
In general, the unique characteristics of MU are shared among current and adjacent cells, thus, there will be no need for re-authentication when passing through heterogeneous cells. As to direction and speed, the SDN controller can recognize the next cell, which in turn informs the BC. The BC checks the \textit{sec\_info} unit to ensure that the MU is a trusted cell. A handover process between the two heterogeneous cells for the 5G network is presented in Fig. 3 (Procedure 2). 
\begin{figure}[ht]
	\centering
	\includegraphics[scale=0.050]{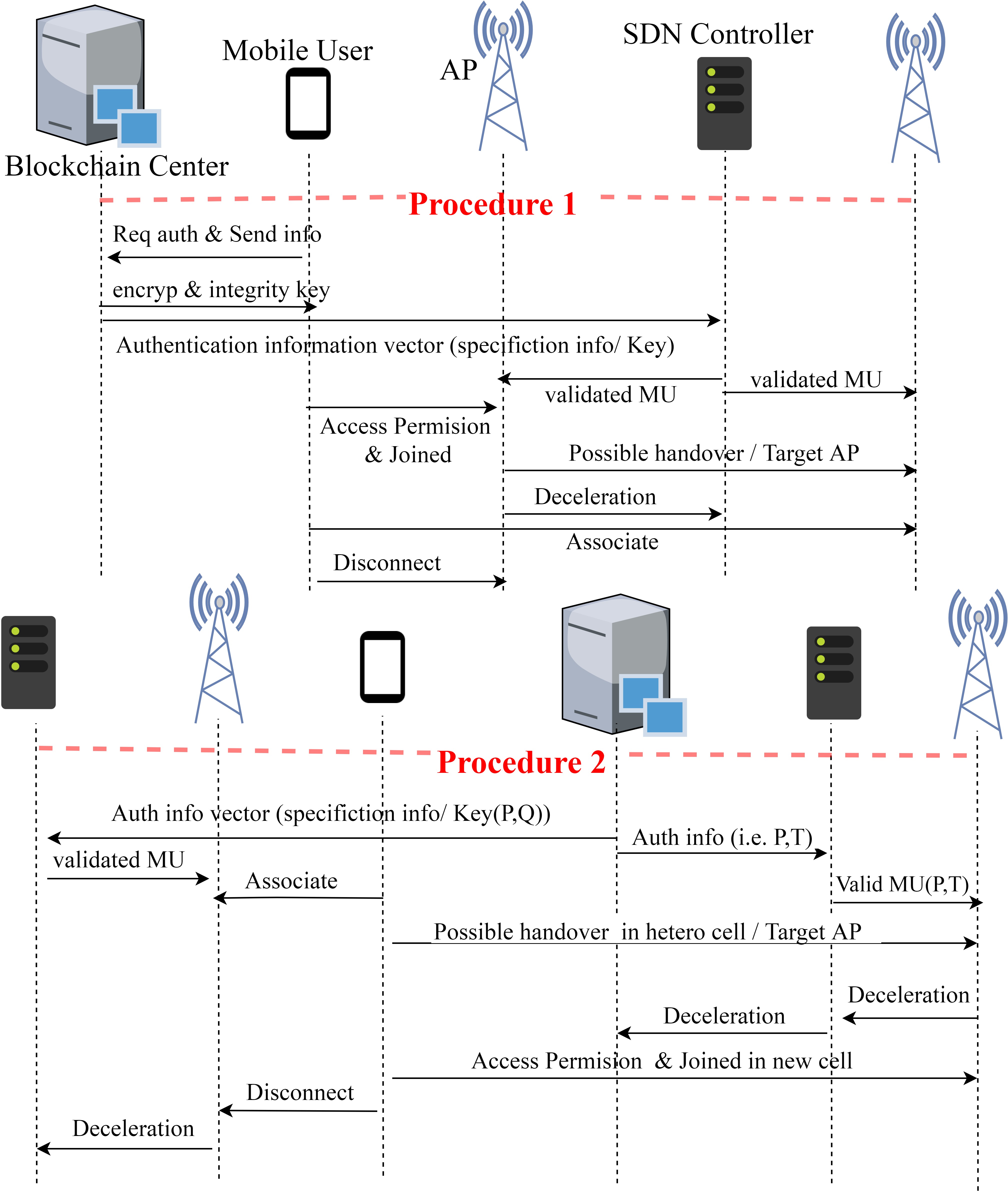}	
	{\scriptsize	\caption{ {\small Registration procedures and behavior in a cell} }}%
\end{figure}
After registering the MU, BC sends its information to the SDN controller of the cell and its adjacent cells, which can be heterogeneous. The SDN controllers in each cell run MU validation operations on APs. The MU is registered in the available cell and intends to go to the neighboring heterogeneous cell that sends the request to the AP of the same cell. The AP then communicates to the controller and accepts the MU request as BC has already considered it as a valid SDN controller in that cell.\\
The SDN controller informs BC of its position and, after disconnecting from the current AP, the controller of that cell is notified.
The public key \textit{P} is known between the MU and the AP, and it is capable of switching between the APs of a cell. The private key \textit{Q} is applied in signing transactions and decoding data for privacy protection.

An SDN controller in a cell has already validated this public key among AP or BS. The BC sends messages to other adjacent public key cells, to eliminate the need for repeated re-authentication in these heterogeneous cells. This mechanism accelerates the authentication process when passing through few heterogeneous networks and APs, and reduces latency. In general, cells receive sensitive information for this mechanism from \textit{sec\_info}. The \textit{sec\_info}, depending on the speed, the direction of travel and the position considers timeout named \textit{T} and makes the cell controller aware of this. The AP based on this \textit{T} waits for the MU entry or its displacement among the APs. If \textit{T} is over and there is no effect on its request to the AP, there is a probability of compromise behavior of which the BC becomes aware, and this follows the assessment of the status of its transactions and even the possibility of blocking, where this blockages will be reported to other SDN controllers.
Specific user information like IP, physical layer properties, position, velocity, and direction of movement within the cell are collected and the SDN configures the flow controller tables according to the given policy, in a sense that it identifies the location of the next cell controller, which accelerates the authentication process and removes the re-authentication process in entering into other cells, as shown in Fig. 4.
\begin{figure}
	\centering
	\includegraphics[scale=0.049]{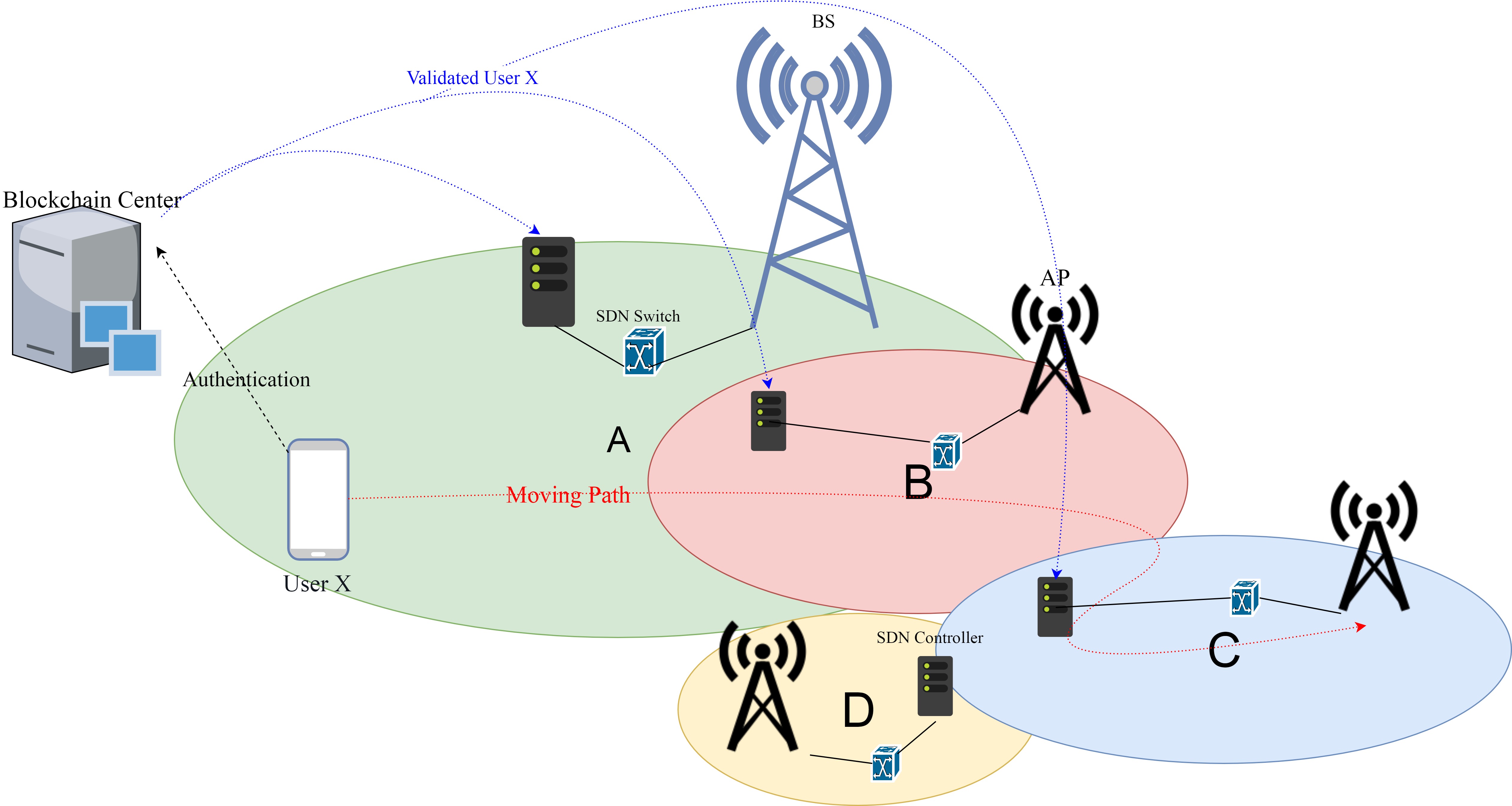}	
	{\scriptsize	\caption{{\small The process of entering and moving among heterogeneous cells in the proposed 5G network structure} }}
\end{figure}
An authentication procedure is applied between the user X and BC, then, user X join the cell A, which is implemented by BC-enabled based on the base SDN structure described in Algorithm 1. Handover authentication does not require any change in user authentication hardware and, similarly, after being registered in BC in other adjacent cells in the direction of the MU, it is run to eliminate the need for repeated authentication when entering cells B and C, which reduces the latency, by removing re-authentication. By predicting the MU route, the SDN controller is ready to serve the MU.
\begin{algorithm}
	\caption{	{\scriptsize User X authentication handover}}
	\label{alg:m1}
	\begin{algorithmic}[1]
		\STATE 
		{\scriptsize	{\textbf{Initial register} : User X // \textit{in  
					Blockchain Center (BC)}} \label{n1}
			\STATE 
			{\quad User X $\to $ \textit{Req authentication and Send info} } 
			\STATE
			{\quad BC $\to $  Send(auth\_User X
				info(spec\_info/ Key))} 
			\STATE
			{\quad User X: receive (encryption \& integrity key)} 
			\STATE
			{\quad BC $\to $Send(User X vector to Predicted Cell)//\textit{SDN controller in near cells} receive info} 
			\STATE
			{\quad SDN Controller: validated User X //\textit{in APs}} 
			\STATE
			{\quad User X $\to $ Associate (State A) } 
			\STATE
			{\quad SDN Controller $\to $Monitor User X} 
			\STATE
			{\quad SDN Controller: Message (BC)// \textit{send info to BC}} 
			\STATE
			{\quad BC: Message (SDN controller Net) //\textit{ send info to SDN controllers}} 
			\STATE
			{\quad\quad \textbf{if} (Mobility or migration)} 
			\STATE
			{\quad\quad\quad \textbf{if} (Possible handover (Target AP)) // \textit{in same cell}} 
			\STATE
			{\quad\quad\quad \quad  SDN Controller: Message (BC)// \textit{send to BC}} 
			\STATE
			{\quad\quad\quad \quad \textbf{Update}(Switch Flow Table)} 
			\STATE
			{\quad\quad\quad \quad User X $\to $Handoff (current AP, Target AP)} 
			\STATE
			{\quad \quad \textbf{else}} 
			\STATE
			{\quad \quad \quad \textbf{SDN Controller}: Message (BC)// \textit{send to BC}}  
			\STATE	{   \quad \quad \quad \textbf{Update}(Switch Flow Table)} 
			\STATE	
			{\quad \quad \quad User X $\to $ Handover (current AP, Target AP) //\textit{ in other cell}} 
			\STATE
			{\textbf{}end}}
	\end{algorithmic}
\end{algorithm}
\subsection{Managing BC Keys in Handover}
The structure and content of BC provide an approach to managing the key in the 5G heterogeneous cells and networks, which in turn reduces key transfer time among cells for users’ handover. The focus here is on the BC on the management of keys in heterogeneous cells to achieve a scalable and light-weight transfer mechanism through the BC. The duty of the BC is to remove a third party (intermediary) in transactions. Key transportation handshake can be eliminated by applying the BC mining method, that is, the messages are approved by the BC instead of the third party. 

The BC structure the excess additional units in the handshake process for validating and authenticating the authenticity of the previous and traditional methods. The handshake process according our approach is shown in Fig. 5.
\begin{figure}
	\centering
	\includegraphics[scale=0.49]{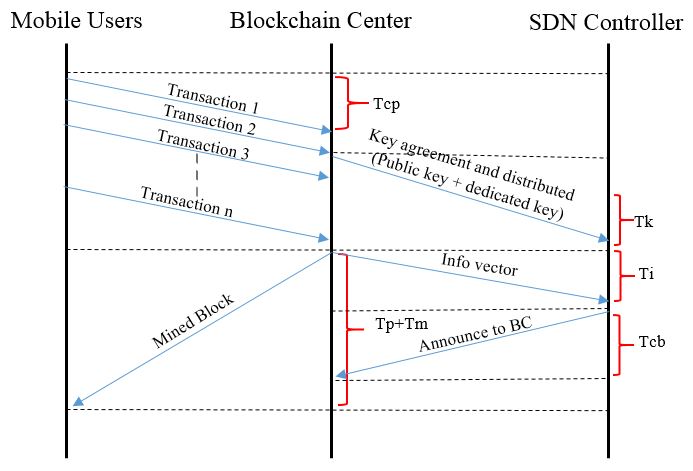}	
	
	{\scriptsize	\caption{{\small The key transmission for BC structure} }}
	
\end{figure}
The collection period (\textit{CP}) allows multiple transactions to be broadcasted to BC, where \textit{Tcp} is the transactions time to BC. Signatures are processed in transactions to assure whether the information in transactions are trusted or not. By applying the public key, the messages are exchanged between the user and BC and only the cryptographic transaction remains to reach the destination and be opened with the user’s private key. As observed in Fig. 5, \textit{Tp + Tm} contain the delay emission time and the mining process. \textit{Tk} is the transmission and distribution time of the public key among controllers and the attributed controller key. 

In Fig. 5, \textit{Ti} is the time to send the user's unique information and features to the controller. \textit{Tcb} is the time to send data from the controller to BC. The key processing time (\textit{Ttotal\_key}) during handover in 5G heterogeneous cells, which includes public/private key emission for the user and the attributed key to the controller in the BC structure to support the handover in heterogeneous cells, is obtained through Eq. (1).
\begin{align}
{T_{total\_key}} = \left[{Nt}\times{Tcp}\right]+Tk+\left({Tp}+{Tm}\right) \tag{1}
\end{align}
where \textit{Nt} is the transaction count obtained through Eq. (2). \textit{Ncp} is the transmitters' theories count \textit{n} is the requisitions count registered in BC at every \textit{m} minute. \textit{Nt} is expressed in Eq. (2).
\begin{align}
Nt = \frac{\left({n}\times{m}\right)} {60s} \times Ncp  \tag{2}
\end{align}
In \textit{Ttotal\_key}, \textit{Ti} and \textit{Tcb} are not contributive in transmitting keys, and are implemented in a co-procedural manner in the key transfer process. Transactions’ time collection is optimized with a minimum of key transfer time.
\subsection{Dynamic Key Management in Handover}
We should be able to manage keys attributes within BC in a way that is compatible with 5G goals by the transaction collection period (\textit{TCP}). Dynamic key management in handover in 5G cell is achieved by using our dynamic \textit{TCP}. To decrement the side effect of variables, the approach of detection variable is engaged in our scheme.

We consider 1ms for pattern metric to measure the efficiency of different collection periods. Thus \textit{Ttotal-All} is a sum up number of transactions in BC. \textit{Nt-1} is the average processing time in 1ms under different collection periods. Basing on the Equation (1) and (2), we can derive the number of transactions on \textit{n} cells as Equation (3).
\begin{align}
{T_{total\_All}} = \left[\left({Nt}\times{Tcp}\right)+Tk+\left({Tp}+{Tm}\right)\right] \times n \tag{3}
\end{align}
Estimated key transfer time is calculated apply different collection periods. The optimized \textit{TCP} time is elected according to the minimum key transfer time. The \textit{TCP} is presented in Algorithm 2, where it is applied in the BC for key transfer.
\begin{algorithm}
		\caption{ 	{\scriptsize Optimize TCP}}
	\label{alg:m2}
	\begin{algorithmic}[1]
		{\scriptsize
			\STATE
			{BC (\textbf{Received} (TCPi)) \textit{// TCP1, TCP2….. TCPn} } 
			\STATE
			{BC = \textbf{Accounting}(mining) } 
			\STATE
			{BC $\to $ \textbf{initialize} ( SDN Controller ) } 
			\STATE
			{SDN controller$\to $ Announce to BC// \textit{Path traffic }} 
			\STATE
			{\quad For (i=0; i<=n ; i++)//\textit{ n is number of transactions}} 
			\STATE
			\quad\quad	{Call \textbf{Equation()}// Calculate the number of transactions  (Equation 3)} 
			\STATE
			{\quad\quad\textbf{Calculate} (Ti)} 
			\STATE
			{	\quad\quad Call \textbf{Traffic cell} (Pi)} 
			\STATE
			{ 	\quad\quad Tcp[i]$\to $Tcp } 
			\STATE
			{End for} 
			\STATE
			{$Tcp^M$ = min (Tcp)} 
			\STATE
			{Return $Tcp^M$} 
			\STATE
			{end} 
		}
	\end{algorithmic}
\end{algorithm}

\section{Efficient Privacy protection in SDN based 5G network}
Data privacy implies the right to separate network users from threats and retaliations against their data. By reducing the size of the 5G heterogeneous cells and networks, users may move among many cells before the communication session is completed, and may be trapped in non-trust APs or become compromised during handover.
Privacy protection is a challenge in 5G. The available approaches \cite{m5,m17} for privacy protection apply a complex key agreement, mutual interaction, by adding signs to protect data, which could lead to latency, and introduce computational load encryption and complexity to the AP \cite{m18}, something not appropriate for small and low power 5G cells. The multiple requirements and the existence of the third party as are considered as a bottleneck in privacy protection, not fit for 5G. By applying the encrypted keys in BC during the communication process, if reading a record is sought, the private key associated with it should be known. In any situation where the attackers do not have the key, and what they get is useless.

In our approach, SDN controllers can select multiple paths for transmitting different parts of the data stream, according to heterogeneous network coverage. Based on the applications, some of the network paths are more sensitive and should be selected. As long as the user is authenticated and be in the network’s coverage, the stream is routed through the controller of that cell, and decrypts the user data based on his private key, and then reorganizes the stream received from multiple paths.
The proposed approach is capable of determining free traffic and network paths through the SDN controllers, something suitable for the 5G network. By choosing multiple routes between APs or femtocells in data transfer, the traffic on the 5G network would be reduced. 
The privacy protection mechanism where SDN and BC are presented in Algorithm 3. 
\begin{algorithm} 
	\caption{ {\scriptsize Blockchain-enabled Privacy Protection using SDN in 5G}}
	\label{alg:m3}
	\begin{algorithmic}[1]
		\STATE	{\scriptsize
			{User A $\to $  Announce to SDN controller// U\textit{ser A wants to send information to User B} }
			\STATE
			{\quad \textbf{If} (user A== authentic in BC \&\& trust in cell)}
			\STATE
			{\quad\quad SDN controller $\to $ Announce to other  }
			\STATE
			{\quad\quad SDN controller ( Check other traffic cells)}
			\STATE
			{\quad\quad SDN controller  $\to $ \textbf{calculate} (optimize (path))}
			\STATE
			{\quad\quad SDN switch $\to $ \textbf{update flow table()} }
			\STATE
			{\quad\quad User A = \textbf{Allocate} (WK) // \textit{allocate bandwidth to user form SDN controller}}
			\STATE
			{\quad\quad User A$\to $ \textbf{Encrypt} (send data (dK)) // \textit{encrypt and send data parts to SDN switch through AP}}
			\STATE
			{\quad\quad VK= \textbf{min} (Ts, tr) // \textit{size in bytes to be transferred with TS}}
			\STATE
			
			{\quad\quad SDN switch$\to $ forward data//\textit{ base policy} }
			\STATE
			{\quad\quad SDN Controller =\textbf{} monitor\-data\-transient(User B)}
			\STATE
			{ \quad\quad  User B $\to $ \textbf{Received} (data)}
			\STATE
			{\quad\quad User B = \textbf{Decrypt} (data) //\textit{ using private key and re-organize data}}
			\STATE
			{else}
			\STATE
			{\quad\quad \textbf{Add to block}()}
			\STATE
			{end}}
		
	\end{algorithmic}
\end{algorithm}
In this algorithm, \textit{K} is the count of the network paths that the SDN controller selects for data transfer. The symbol \textit{dk} is a different data section that can be sent in \textit{K} directions in a simultaneous manner. The symbol \textit{tr} is the time of data transfer inside the involved cells. \textit{Ts} is the delay threshold for 5G applications. For example, some services like transferring email that can withstand a long delay or online games with a slight threshold of delay before \textit{Ts} is required. \textit{WK} is the bandwidth allocated by the SDN controller is in accordance with the traffic situation of different networks, and \textit{VK} is the transmitted data volume in multiple paths with the delay of the delay threshold application. The count of \textit{K} paths here is through the balance between the privacy level and the complexity of the adjustable system through the SDN controller. Privacy protection for users is programmable through the SDN controller, and the advantages of BC are essential for future requirements.
\section{Consensus algorithm used in our BC }
In order to verify and record every transaction in open blockchains, it has to go through consensus process. POW \cite{m18} has been mainly the consensus protocol used in a many decentralized platforms including Bitcoin and Ethereum, which could be problematic in 5G environments due its high computational power and latency \cite{m15,m16}. 

Also, the original DPOS has the limitations of vulnerability to centralization as the number of evidences is limited, and being exposed to fault of real-life voting.
To address this issues in our approach, we have solved the DPOS limitations \cite{m16} in our upgraded DPOS algorithm with the help of SDN controllers in any cells and the network between them in the our proposed 5G architecture.

Our mechanism acts like a board of directors where SDN controllers in the cells vote on a number of SDN controller candidates, in a sense that they would be responsible for verification and billing. In general, for constructing each block, SDN controllers are applied for electing representatives who would produce blocks based on collaboration. Representatives monitor each other’s performance, and when one is out of line is either omitted or does not get votes. DPOS can reduce the count of the nodes associated with the authentication and accounting process in a significant manner. 

In our context, the SDN controllers perform transaction processing and the addition of a new block to the BC. When the SDN controllers’ miners succeed, they can receive a certain volume of information of their interest as a reward, indicating that the miner is the processor of the information he/she is interested in. Miners are nodes that control a large amount of information on the network, because in each cell the controller monitors users. It is possible that someone else would be interested in this information, in this situation, the user can request from the miner for this information, in a sense that the miner verifies the access control policy in the BC and then shares the data. Here, the advantage is that the user will delete the decryption process after the verification. The benefits of DPOS include cheap, scalable, energy-efficient transactions. A partial centralization in devising creation of the blocks makes this algorithms’ functionality better than its counterparts. Devising a new block in Bitcoin takes only 10 minutes, while the EOS, by applying DPOS, do the same in less than one second \cite{m16}. 
Algorithm 4 shows the mining procedure in our approach. 
\begin{algorithm}
	\caption{  {\scriptsize upgraded DPOS  Algorithm}}
	\label{	alg:m4}
	\begin{algorithmic}[1]
		{\scriptsize
			\STATE
			{BC = \textbf{Get\_info} (H, VB, HPre, tstamp, S, Trans) //\textit{ BC strat updating and get require info like, ( H: Block Header, VB: Block Version, HPre: Previous Block Hash, tstamp: Timestamp, S:  number of SDN controllers, Trans: transactions Trans = [T1; T2:::Tn] )}} 
			\STATE
			{Group\_Agent = \textbf{Voting}() // \textit{select group of agents for mining}} 
			\STATE
			{BC $\to $  \textbf{Announce} (Candidate mining) // \textit{sending require info to agent}} 
			\STATE
			{Initialize bool variable \textit{K} = False} 
			\STATE
			{Integer P = 0; } 
			\STATE
			{\quad \textbf{While} (NOR K) \textbf{do}{} 
				\STATE
				{\quad \quad  transaction order = \textbf{random}(n) // \textit{range [1:n]}} 
				\STATE
				{\quad \quad \textbf{Calculate} ( Merkle tree root (Root\_P))} 
				\STATE
				{\quad \quad Root\_P$\to $ basing on the transactions in payload} 
				\STATE
				{\quad \quad \textbf{Create hashed block header} ( Hn = VB$\| $HPre$ \|$ tstamp$ \|$ S $\|$ Trans)} 
				\STATE
				{       \quad \quad \quad    \textbf{while} (NOR K \& NOT got DPOS) \textbf{do}} 
				\STATE
				{\quad \quad \quad \quad Group\_Agent$\to $ \textbf{mining}(block);}
				\STATE
				{       \quad \quad \quad \quad    Create header: Hverify= VB$\| $HPre$ \|$tstamp$\|$S$\|$Trans;} 
				\STATE
				{\quad \quad \quad \quad The string to Hn= Hverify$\|$nonce; } 
				\STATE
				{    \quad \quad \quad \quad      Result = hash(Hn $\|$ nonce)} 
				\STATE
				{\quad \quad \quad \quad\textbf{Cooperate} (Group\_Agent)} 
				\STATE
				{\quad \quad \quad \quad Extract  nonce = \textbf{getNonce}(Hn);} 
				\STATE
				{\quad \quad \quad \quad RootP basing on the transactions in Bpayload}
				\STATE
				{\quad \quad \quad \quad nonce ++ ;} 
				\STATE
				{\quad \quad \quad \quad \quad \textbf{if} (NOT got DPOS) \textbf{then}} 
				\STATE
				{\quad \quad \quad \quad \quad Group\_Agent$\to $ mining(block);} 
				\STATE
				{\quad \quad \quad \quad \quad  P= (nonce - 1) into nonce field} 
				\STATE
				{\quad \quad \quad \quad \quad Return (nonce - 1);} 
				\STATE
				{   \quad \quad \quad \quad \quad      K= true;}
				\STATE
				{               \quad \quad \quad \quad \quad   \textbf{ else if} (receive DPOS) \textbf{then}} 
				\STATE
				{\quad \quad \quad \quad \quad  Group\_Agent$\to $ \textbf{mining}(block);} 
				\STATE
				{\quad \quad \quad \quad \quad  K= true;} 
				\STATE
				{        \quad \quad \quad \quad                return NULL;} 
				\STATE
				{         \quad \quad \quad \quad         \textbf{   end if}} 
				\STATE
				{         \quad \quad \quad     \textbf{  end while}}
				\STATE
				{\quad \quad \quad \textbf{}end while}} 
			\STATE
			{ \quad \quad \quad  \textbf{If} (Miner == success) \textbf{then}{} 
				\STATE
				{\quad \quad Miner (i) = \textbf{Resive\_data\_interest} ();} 
				\STATE
				{        \quad \quad    \textbf{Share\_data} (Miner(i));} 
				\STATE
				{$\}$\quad$\}$} 
				\STATE
				{end} }
		}
	\end{algorithmic}
\end{algorithm}
\section{	Evaluation and Results}
Experiments are run to evaluate the delay, efficiency, and comparability of the proposed approach. The results are obtained through OMNeT ++ 5.1 \cite{m20} with the INET 3.6.4 framework. The INET framework, which is involved in the implementation of the SDN switch and controller, can support the SDN\cite{m33} and the BC function \cite{m21}. The BC encompasses the consensus algorithm (DPOS) for users in the 5G network. We define certain categories of messages among BC and cells to help achieve a common view of the blockchain among all participating MU.
\begin{itemize}
	\item \textit{Ini\_reg (net id, Loc, Dire, from, phys\_ly, Rtt, to), Ini\_reg \_ack (peer list):} Allows the users to discover BC and broadcast their information among near cells.
	\item \textit{Get\_block \_list ()}: Request from BC, the list of blocks available with number of SDN controllers for mining.
	
	\item 	\textit{Get\_tran\_list()}: Request for transactions in the cells (not yet mined into a block).
	
	\item \textit{	block(block\_header, tran\_list)}
	
	\item 	\textit{block header(hash, timestamp, miner, merkle root)}
	
	\item 	\textit{tran(in\_list, out\_list)}: Transactions have a list of inputs in each cell which it is spending, and a list of outputs which it creates in that cell.
	
\end{itemize}
In order to warrant that all MU have a uniform view of the blockchain, we define and consider the rules followed by the BC so as to get consensus. We define a MU in the 5G cell by the tuple \textit{(Ini\_reg, P, Q, T, firmware)}, where \textit{(P, Q)} is the public and private key pair for the MU, firmware is a function use defined algorithm in our approach. To evaluate the feasibility and performance of our proposed model, we define the following metrics to measure the feasibility of our proposed model:
\begin{itemize}
	\item 	\textit{Ttran} : Transactions added to the blockchain. 
	\item	\textit{Tblock} : Blocks added to the blockchain per second by SDN controllers via DPOS algorithm.
\end{itemize}
To evaluate the comparability of the proposed approach among heterogeneous 5G cells, a network containing 30 heterogeneous cells with a distance of 200 meters between the two APs in two cells and the MU with 5KM /h speed change and direction of every 3 seconds were selected. In each cell, the controller updates the flow tables of each cell, based on the parameters of the users who register in BC. The details of the simulations are presented in Table I.
\begin{table}[H]
	\begin{center}
		\caption {{\scriptsize Stimulation parameters}}
		\begin{tabular}{|c|c|}
			\hline
			
			{\tiny	{\textbf{Simulation Parameters}}} & 	{\tiny{\textbf{Values}}  }\\ 
			\hline 
			{\tiny	Simulator}	& 	{\tiny	OMNeT++ 5.1 with INET 3.6.4 }\\ 
			\hline 
			{\tiny		Number of Cell	}& 	{\tiny30	}   \\ 
			\hline 
			{\tiny	Number of request/response to BC}	&	{\tiny 2000  }\\ 
			\hline 
			{\tiny	Number of Transactions}	&	{\tiny 1200   } \\ 
			\hline 
			{\tiny	Number of SDN Switch	}	& 	{\tiny90   } \\ 
			\hline
			{\tiny		Number of SDN controllers	}& 	{\tiny30 }   \\ 	\hline 
			{\tiny	Cell radius}		& 	{\tiny100 m  }  \\ 
			\hline
			{\tiny	User mobility speed	}&	{\tiny 5 Km/h }   \\ 
			\hline 
			{\tiny	User mobility direction	}	&	{\tiny Random    }\\ 
			\hline 
			{\tiny	Total number of users	}	&	{\tiny 600   }\\ 
			
			\hline 
			{\tiny	distance between two AP		}&	{\tiny 200  }  \\ 
			
			\hline
			{\tiny	Length of packets registered in BC (M)		}&	{\tiny 32 byte }  \\ 
			
			\hline
			{\tiny  Length of packets from BC to the controller(N) 	}&	{\tiny  16 byte}   \\ 
			
			\hline	
			{\tiny Transmit power (Pt)		}&	{\tiny 1726 mW  }  \\	
			\hline	
			{\tiny  Receiving power (Rx)}&	{\tiny  1340 mW }  \\	
			\hline	
			{\tiny  The influence of the number of miners 	}&	{\tiny  $ N=\left({\mu }_{t}=1500,{\sigma }^{2}=4\right) $   } \\	
			\hline	
			{\tiny  Block Size		}&	{\tiny  4 byte } \\	
			\hline	
			{\tiny  Transaction Counter		}&	{\tiny 1-9 byte }  \\	
			\hline	
			{\tiny  Block Header		}&	{\tiny  80 byte }  \\	
			\hline	
			{\tiny  Prev\_block\_hash		}&	{\tiny  32 byte}  \\	
			\hline		
		\end{tabular}
	\end{center}
\end{table}
\textbf{Signaling overhead}: it contains a pattern or additional information to enhance performance of the wireless communications. It is related to registering the MU in BC. This is to be compared with the network-based signaling overhead and POW-based models, which use POW \cite{m22}. This method is similar to that of \cite{m22}. Here, Eq. (4) is applied for this analysis.
\begin{align}
{Signaling overhead (Sover)} =\frac{ \left({B}\times{M}\right) }{ t} + \frac{(N)}{ t }\tag{4}
\end{align}
where \textit{t} is the time and \textit{B} is the count of steps between MU and BC. \textit{M} is the length of packets registered in BC and \textit{N} is the length of packets sent from BC to the controller. As results indicated in Fig 6, our proposed approach has less overhead than the network-based and POW-based models because the DPOS algorithm was applied in our work together with the SDN controllers for managing each cell.

In the POW-based model where POW is applied it has extra overhead and does not allow the SDN controllers to be applied in mining the handover. The network based method requires the third party in communications and a variety of authentication servers among heterogeneous cells in a 5G network. As observed in Fig. 6, an increase in time increases the signaling overhead because there are more requests for registration and authentication among cells. Our approach has less overhead than its counterparts because it is directly registered in the BC and joins the cell with no need to interact with other nodes.
\begin{figure}[H]
	\centering
	\includegraphics[scale=0.38]{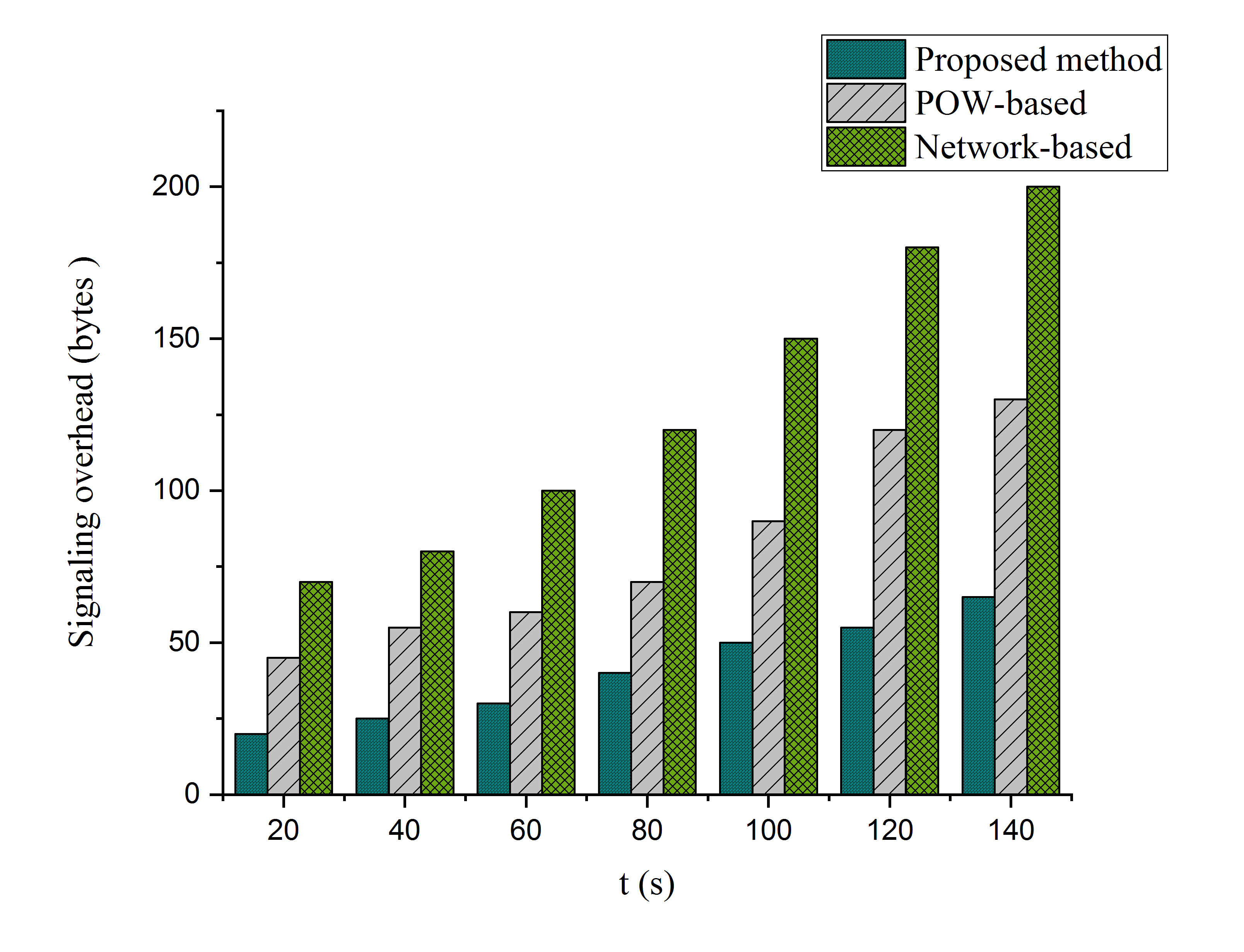}	
	{\small	\caption{ {\small Comparison of signaling overhead of our approach with the network-based and POW-based models} }}
\end{figure}
\textbf{Energy Consumption}: In all subject models, if the transmit rate\textit{ Ctx} is the transmit rate and \textit{Crx} is the received rate, the energy consumption is calculated through Eq. (5). 
\begin{align}
\nonumber E & = \left( Ctx \times Pt \times t1 \right)+ \left( Crx \times Rx \times t1 \right)+
\\ & \left( Pr \times \left( t - t1\right)\right)  \tag{5}
\end{align} 
where \textit{Pt} is the transmit power, \textit{Rx} is receiving power, \textit{t1} is the connection time, and \textit{Pr} is the received power. The energy consumption for the whole network is calculated through Eq. (6), where the signaling overhead must be applied as well: 
\begin{align}
\nonumber En & = \left({n}\times{c}\right) \times \left[\left(\left( \frac{  Sover}{B} \times a1  \right)+a2\right)\right]\\ & \times Ptx \times \left(Ctx \times Crx \right) + \left(Pr \times \left( t-t1\right) \right)   \tag{6}
\end{align}
where \textit{n} is the count of heterogeneous cells, \textit{c} is the count of the controllers and \textit{a1} and \textit{a2} are power constants. The comparison results of the energy consumption of this analysis presented in Fig. 7.
\begin{figure}
	\includegraphics[scale=0.36]{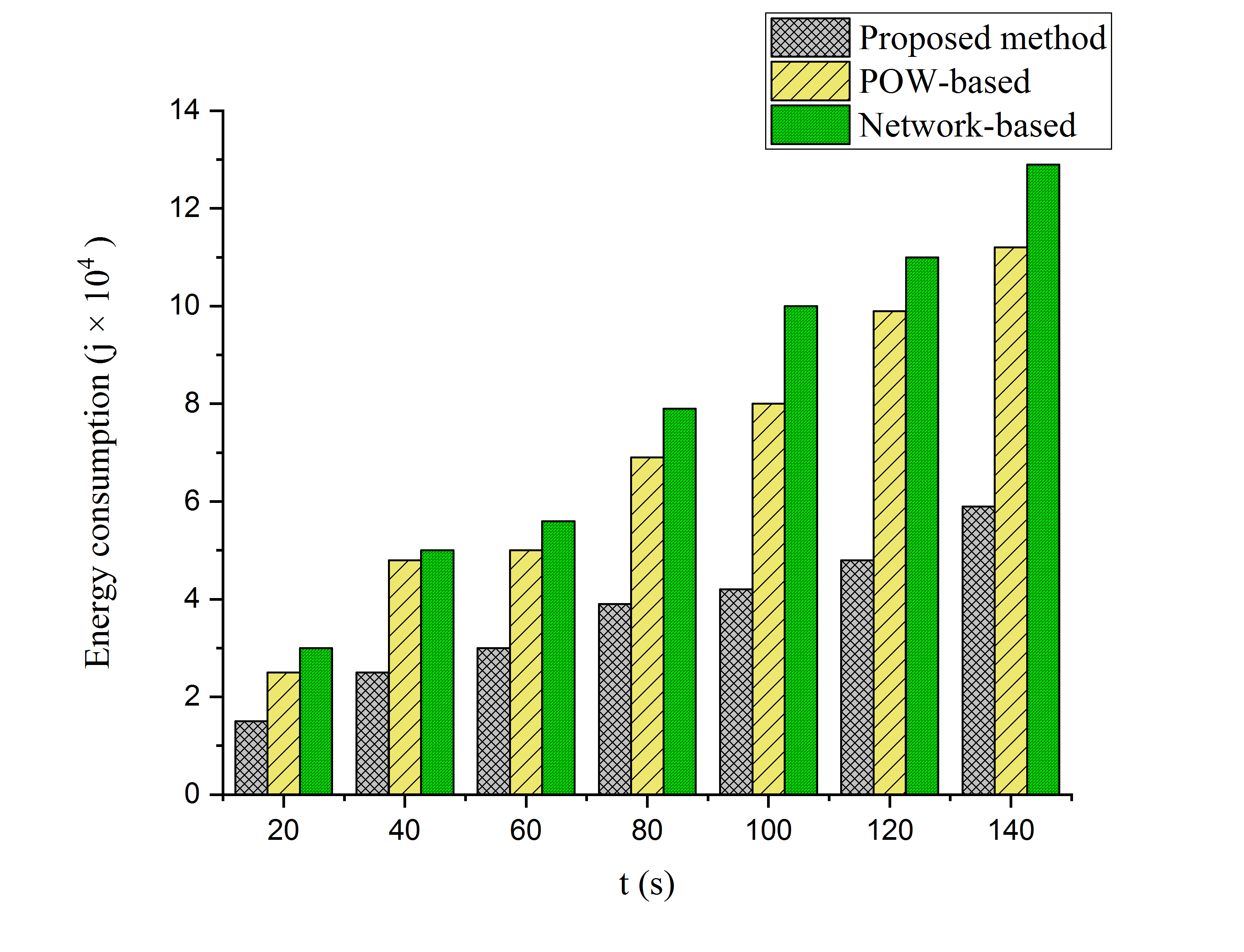}	
	{\scriptsize	\caption{{ Comparison of the energy consumption of our approach with the network-based and POW-based models} }}
\end{figure}%
According to the figure, in the network-based model, the MU reaches to the third part that is responsible for authentication and needs three handshakes in a few steps. In the POW-based, more energy is consumed on POW in comparison with our proposed approach.

\textbf{The influence of the number of miners in our approach}: Here, it is assumed that the mined blocks size through the miners are of the normal distribution, $ N=\left({\mu }_{t},{\sigma }^{2}\right) $  and its function is expressed through Eq. (7).
\begin{equation}
{f}_{\left(x\right)}=\frac{1}{\sigma \sqrt{2\Pi }}\mathrm{exp}\left(-\frac{{\left(x-\mu \right)}^{2}}{2{\sigma }^{2}}\right)         \tag{7}
\end{equation}
The values of the parameters are ${\mu }_{t}= 150$ and $ {\sigma }^{2}= 4$. The effect of increasing the miners’ count on the total service demand indicates the normalized rate of the provided service. These services include consensus-building by miners through the DPOS algorithm and sharing information for users and BC. Fig. 8 shows the results of this analysis.
\begin{figure}[H]
	\centering
	\includegraphics[scale=0.35]{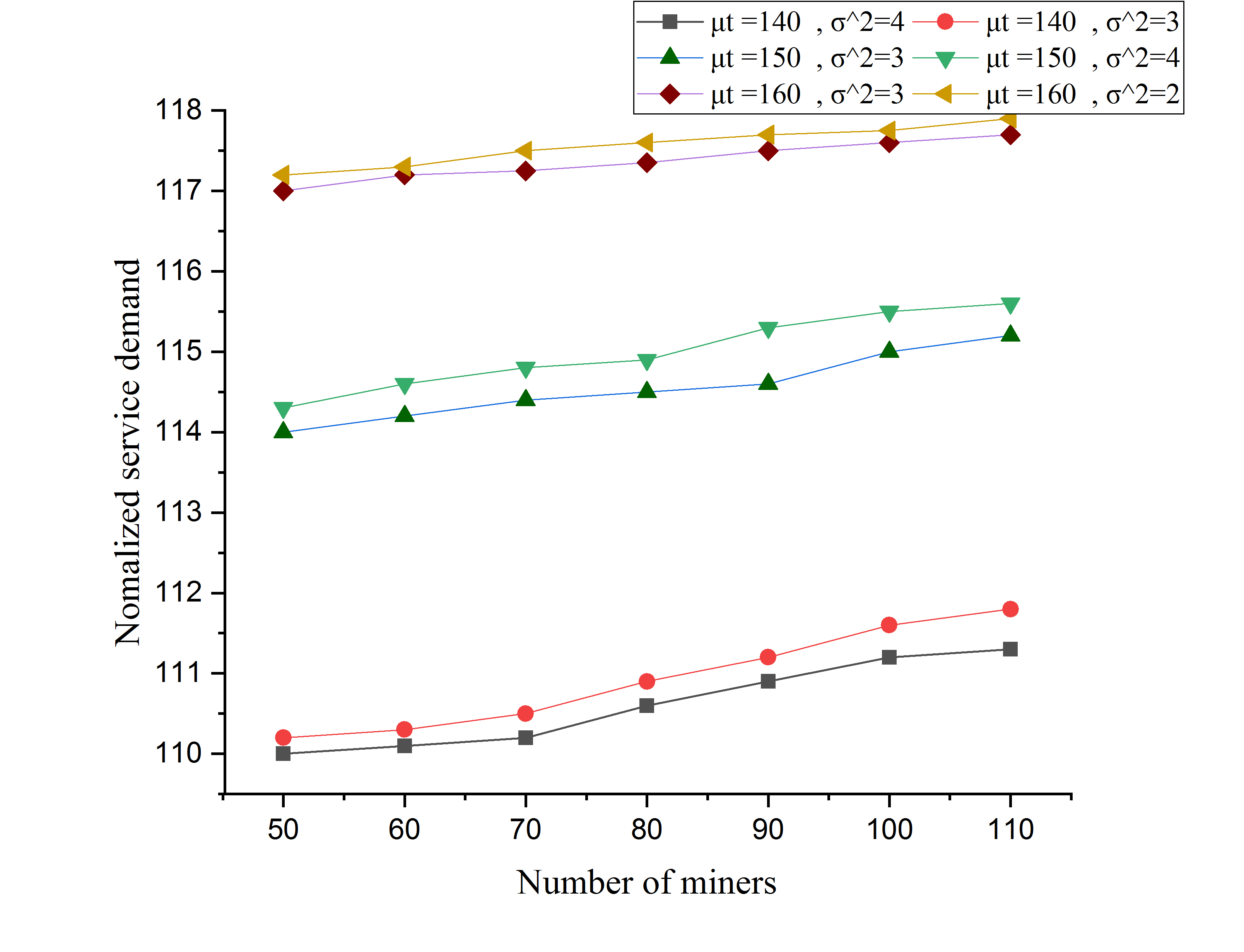}	
	{\scriptsize	\caption{The effect of the miners’ on the offered service rate} }
\end{figure}%
The reason for the increase in the miners’ count is the increase in the heterogeneous cells’ count in the 5G and the possibility of further coverage of the network that is, providing more service. The more the count, the more the miners, indicating that more delegates will collaborate to reach a consensus, thus a reduction in the delays. An increase in  $ {\mu }_{t}$ rate increases the demand for service because this increase in  $ {\mu }_{t}$ increases the average size of a block ending in the miners’ rewards.

\textbf{ Privacy efficiency preserving}:
In this particular analysis, our approach was compared with the two SEMR-ABE \cite{m23} and DACC \cite{m24}, as shown in Table II. \textit{Te} is the time for one exponentiation, and \textit{Tc}{\tiny } is the constant time. \textit{S} is the text code size, and \textit{f} is the code attribute. In our method, something that the user does not need is a cryptographic operation but decryption, thus no overhead as applying the BC properties fits 5G with less delay.
\begin{table}[H]
	\begin{center}
		
		\caption{{\scriptsize Comparison of Transmission Traffic and Calculation Time of our approach with DACC and SEMR-ABE for preserving privacy}}
		
		\begin{tabular}{|c|c|c|c|c|}
			\hline
			{\mbox{	{\tiny\textbf{Scheme}}}} & {\mbox{{\tiny\textbf{Size of Password text}}} }&  {\mbox{{\tiny\textbf{Decryption time}}}}&  {\mbox{{\tiny\textbf{Revocation message}}}}& {\mbox{{\tiny\textbf{	Transfer security}}}} \\ 
			\hline 
			{\tiny	DACC}	& {\tiny	(3f + f)S}&{\tiny f $\times$  Te	} &	{\tiny f $\times$ S}	 &{\tiny no} \\ 
			\hline 
			{\tiny	SEMR-ABE	}& {\tiny	$ {L}^{2}+\left(f \times s\right)+s$	} &{\tiny Te }& {\tiny Te }& {\tiny Yes } \\ 
			\hline 
			{\tiny	Our method }	&{\tiny S } &{\tiny Te }&{\tiny Tc}&{\tiny Yas} \\
			\hline 
			
		\end{tabular}
	\end{center}
\end{table}
The efficiency of Algorithm 3 is measured in two aspects: delay and bandwidth. In case that the source and destination are not for transferring files between adjacent cells, and there are other heterogeneous cells among them, there are K paths via which a SDN controller deals with the file transfer in terms of latency, bandwidth, and data volume. Proposed algorithm has less delay than the network-based (it does not consider traffic and delay threshold). As illustrated in Fig. 9, network-based increases the delay by increasing the data size because the packets are in queue, K represents the number of different paths in our method for sending, and the SDN controller simultaneously sends data from different paths to reduce transmission delay. Also, the effect of this can cause the decrease of consuming bandwidth during transferring packets among cells. As shown in Fig. 10, with the increase of packets transferring at the path between source and destination in a cell, the proposed method is more efficient than network-based as the algorithm considers traffic of another cell and sends packets through efficient path. Here we considered an average of consumption bandwidth in the path with various K=(2,3,4).

\begin{figure}
	\centering
	\includegraphics[scale=0.35]{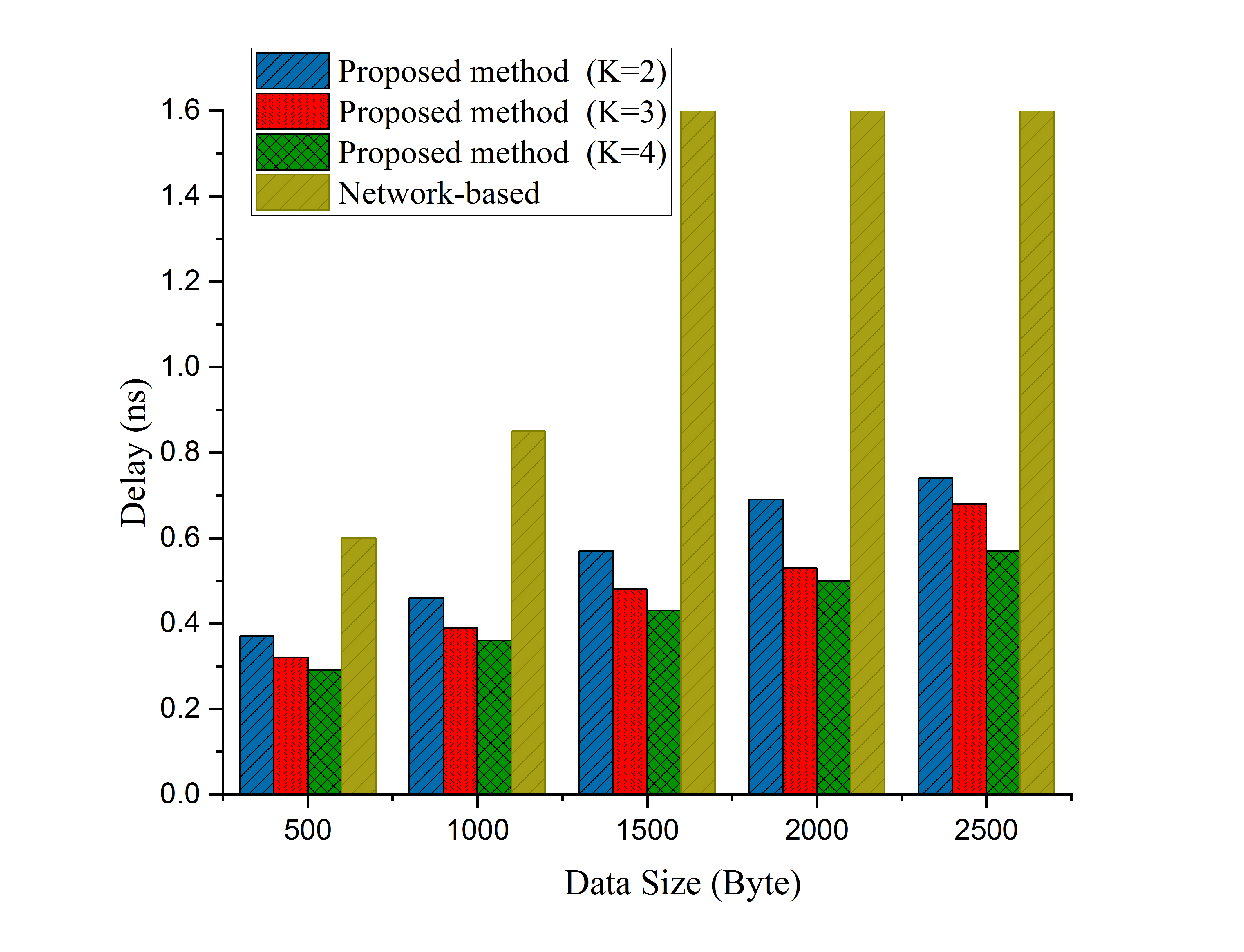}	
	{\scriptsize	\caption{ The efficiency of delay measured between proposed method and network-based model} }
\end{figure}
\begin{figure}
	\centering
	\includegraphics[scale=0.35]{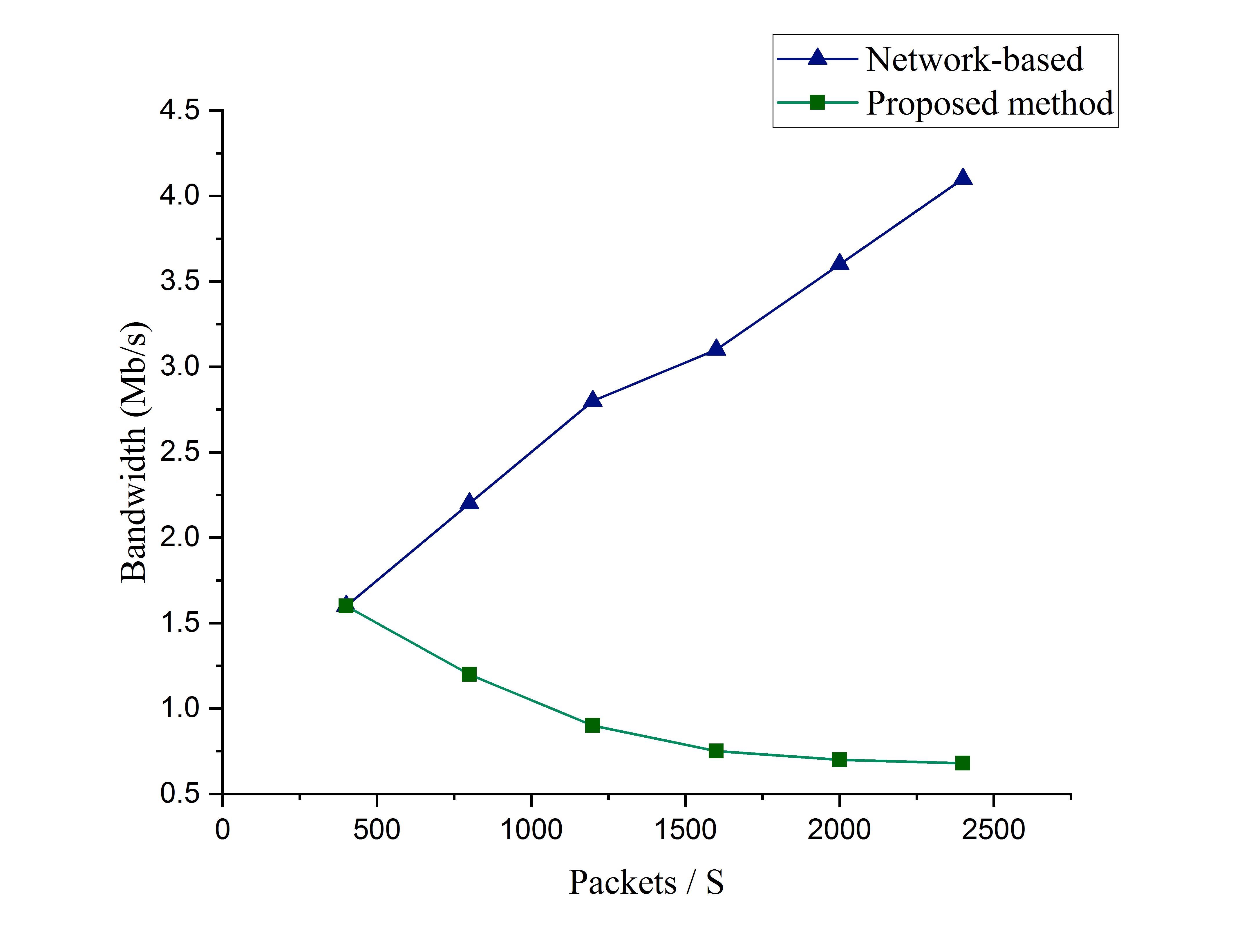}	
	{\scriptsize	\caption{ The efficiency of consumption bandwidth measured in a cell between proposed method and network-based model} }
\end{figure}
\textbf{Handover execution time}: The assessment of authentication handover delays in our approach is compared with both POW-based and network-based models. In POW-based method, users must be registered in the blockchain and upon repeated displacements among the cells become re-authenticated in the cell. These two methods require re-approval and separate protocols among heterogeneous cells for authentication.

In our approach, the user does not need to be re-authenticated when being replaced among the heterogeneous cells because they are valid in adjacent cells and easily handover, removing the re-authentication delay. The comparison between our authentication delay and the other two methods based on the utilization rate in the 5G network is shown in Fig. 11. 
\begin{figure}
	\centering
	\includegraphics[scale=0.37]{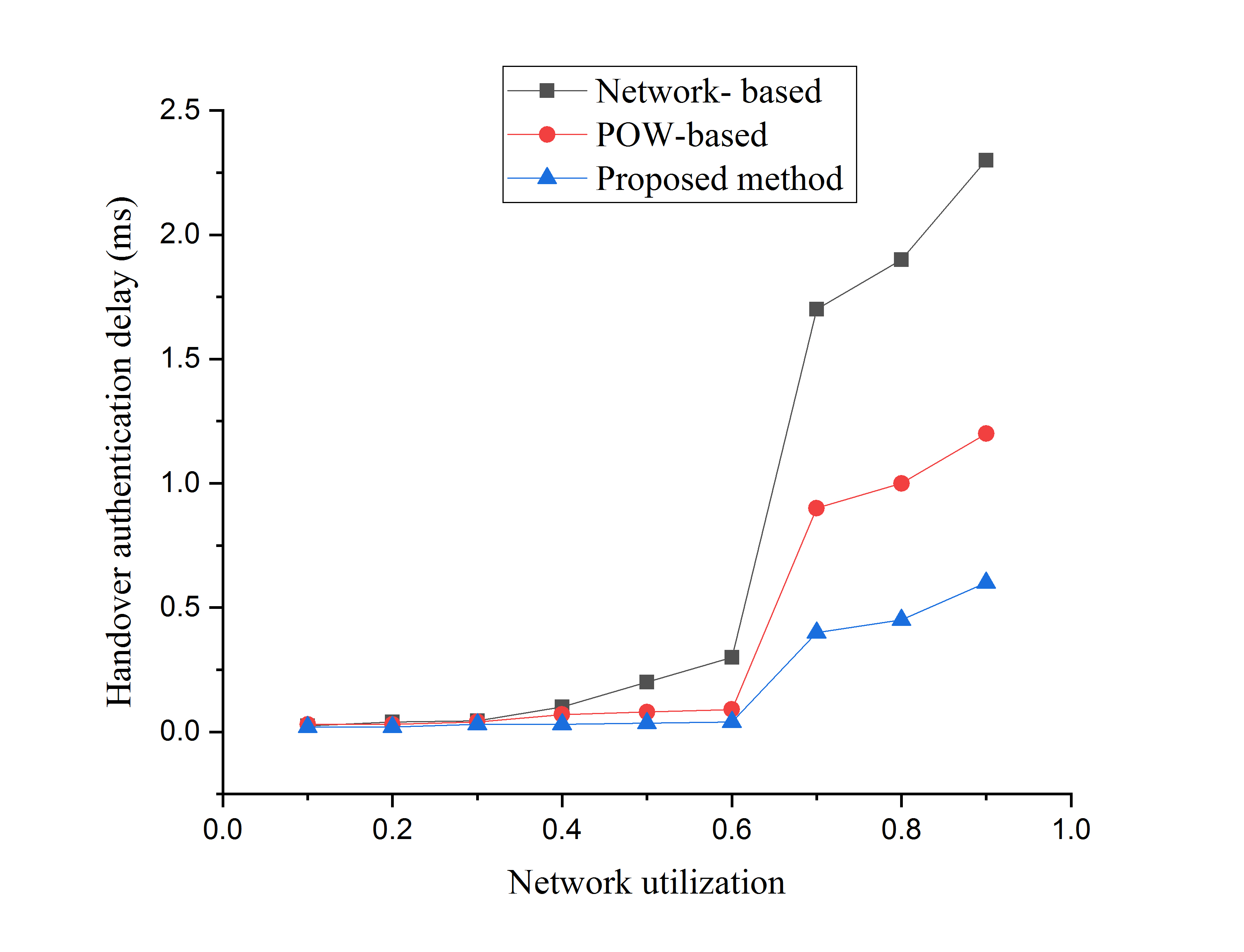}	
	{\scriptsize	\caption{{\small Comparison of handover authentication of our approach with the network-based and POW-based models} }}
\end{figure}
In this context, network efficiency is the total data volume reached the processing rates’ ratio in BC and SDN controllers. The network productivity rate is defined as the different load conditions in the network. When the network load is low, the authentication delay does not have a problem in our approach and other methods. But, when the count of the users increases and there is mobility among the cells, and the data transmission operation is run, the network load increases in the other two method as opposed to our approach, which resulted in an delay less than 1ms making it suitable for 5G.

\textbf{Processing Time of Cryptographic}: In this analysis, the time spent on cryptographic is assessed with the objective to apply Eq. (1) and algorithm 2. For this purpose, the key transfer time must consider the cryptographic approach. The efficiency of the cryptographic approach where the key transfer procedure was taken into account. Except for the mining process, an increase in transactions’ count, the processing time increases in a linear manner. The mining algorithm is always a mined header, containing multiple transactions. The processing time of mining is the mean value of multiple simulations. The processing time of DPOS and POW are shown in Fig. 12. Due to the cooperation of representatives (30 representatives here), DPOS reaches a consensus in less time.
\begin{figure}
	\centering
	\includegraphics[scale=0.37]{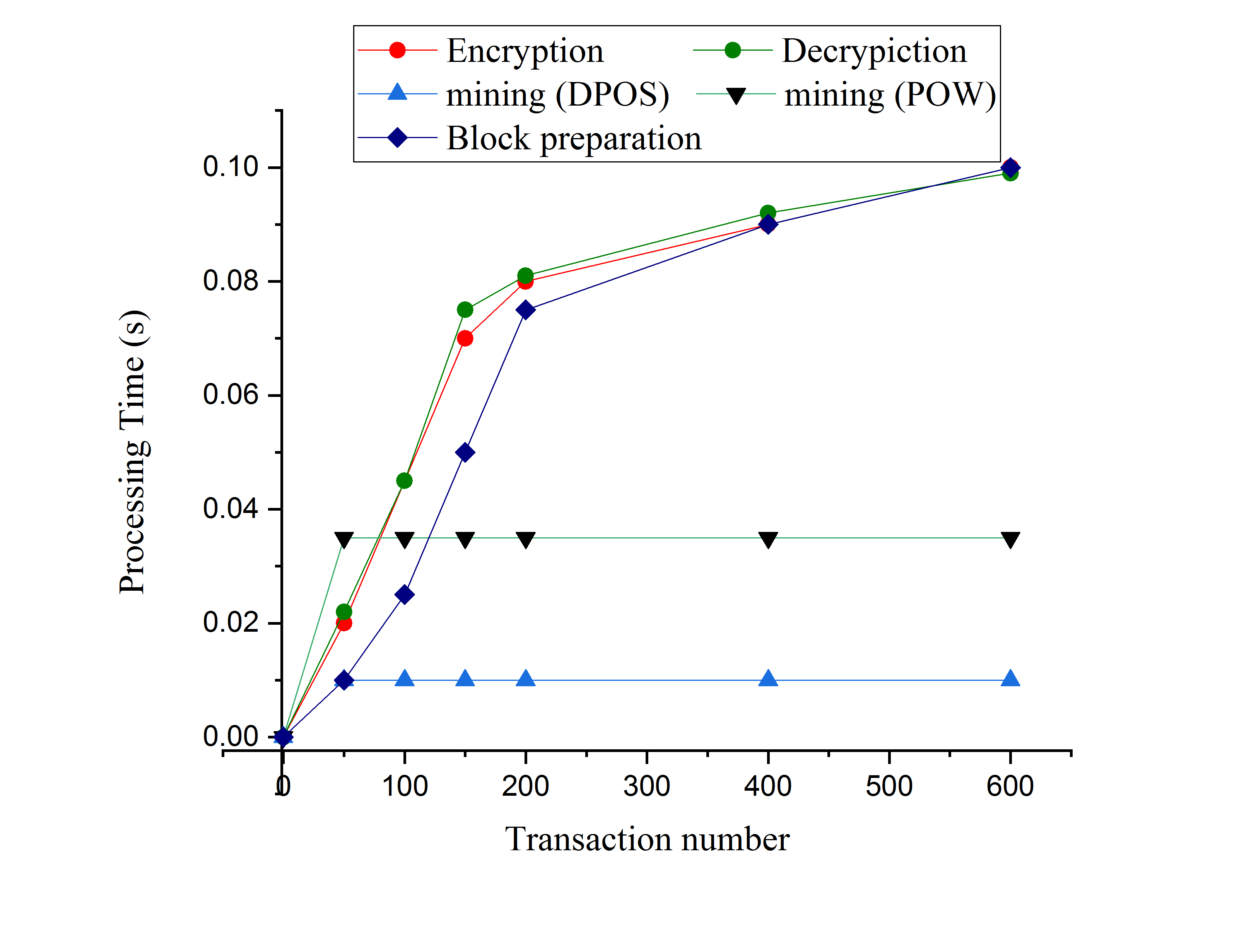}	
	{\tiny \caption{{\small Cryptographic time-out assessment through a key transfer procedure} }}
\end{figure}

\textbf{ Attack model}:We have considered two classes of attacks.

\begin{itemize}
	\item\textit{ 	Class 1 attacks:} DOS/DDOS attacks like, UDP Flood, SYN Flood, TCP Flood which are common in the network level \cite{33}.
	\item \textit{	Class 2 attacks:} Those attacks that can occur when MU want to join cell or BC like, ID spoofing, Authentication attack, Link-ability attack and Numb attack\cite{34}.

\end{itemize}
Attacks in class 1 usually execute by MU in cells after joining and registering in the network for running down infrastructure in the cells. BC and SDN controllers usually face with attacks in class 2 since these attacks work with BC and SDN controller directly. For example in Authentication attack, MU frequently wants to join BC or other cells with malicious behavior, or wants to join network with fake or block ID for communication with other MUs.
	For assessing the attack detection capability of the proposed architecture, in the simulation we assumed that in 20 cells there were 400 MUs, 50 of which were malicious (i.e. occupying bandwidth by sending duplicate packets) and 100 AP (20 under compromise and file transfer barriers). In our scenario, if AP is compromised and the MU attacks (i.e. class 1 or class 2 attacks) inside the cell, the SDN controller is able to detect it based on traffic generated, packet type, and consumed bandwidth. The SDN controller is also able to detect the attack and notify BC and other SDN controllers based on the patterns defined by the BC or mobile operator. Also, the BC has valid ID MU (specified by mobile operators) associated with the registration which then can be used to detect malicious MU. The probability of detecting an attack in the proposed architecture during 100 times of implementation of this scenario is shown in Fig. 13. The results shows that the proposed approach easily detects attacks and blocks  ID them by the SDN controllers and BC.

\begin{figure} 
	\centering
	\includegraphics[scale=0.35]{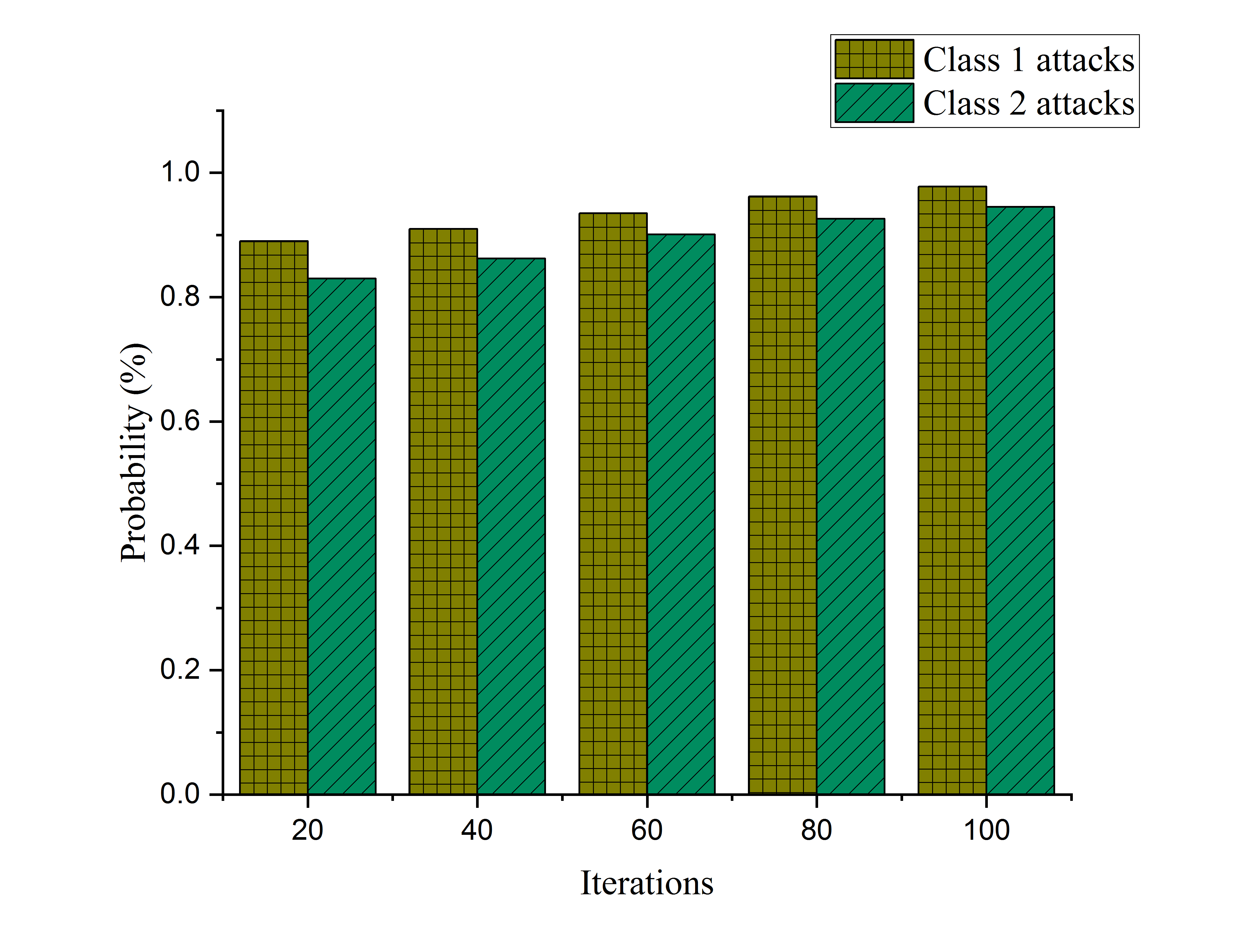}	
	{\scriptsize	\caption{ Percentage of attack detection probability in the proposed architecture} }
\end{figure}

\section{Related Work}
The 5G network requires high capacity and efficient security mechanisms to support data traffic\cite{m1, m25}. The concentration of heterogeneous networks and vast expansion of small base stations have led to the selection of the 5G network \cite{m25}.Many applications supported in 5G require high privacy and reliability against malicious attacks, like mobile banking and social network applications\cite{m19}. The common methods for secure communications in 3G and subsequent wireless networks are based on the controllability and cryptographic modifications that impose constraints and challenges for 5G\cite{m5,m25}.

To support the increased data traffic, 5G networks require high capacity together with strong security mechanisms. With the advent of new communication models in 5G in heterogeneous environments like vehicles\cite{m1}, SDNs\cite{m5}, IoTs\cite{m6}, and fog computing\cite{m1,m17}. The common approaches to having a secure connection in 3G and subsequent generations are based on cryptographic exchange control\cite{m18}, which requires different authentication servers and protocols to support different networks and channels. When the authentication servers are located in another location and are remotely accessible, security challenges rise which are not suitable for 5G \cite{m5}. According to \cite{m16}, due to the frequent displacements among cells, the existence of repetitive authentication is inevitable which may take one-hundredths of milliseconds to identify users, something that would not be acceptable for 5G communications.


In \cite{m27}, a 3rd Generation Partnership Project (3GPP) provided a specific presentation of hierarchical keys and the flow of handover messages for animated scenarios. Although there exist keys for the handovers, and different handovers are required for different scenarios, this approach increases the delay and the handover’s complexity when entering into different 5G cells.

An authentication handover approach was developed by \cite{m28}, including direct authentication between the user and the AP based on public cryptography. Their proposed mutual authentication and the key agreement does run authentication according to new networks through three-way handshake without having contact with a third part, like authorized and server. Although this authentication handover procedure seems simple, it increases the cost of computing and delays due to the overhead it imposes for exchanging encryption over the wireless mediators. Accordingly, the transfer of a digital signature for the 5G wireless networks is not effective in their approach because of exchanging more cryptographic in overhead.

The privacy presentation for SDN / NFV base architecture in 5G was discussed in \cite{m29} with a special focus on network architectures where the core of their mobile packets were for the SDN/NFV structure. This study was focus on  the privacy in 5G scenarios like position protection, identity protection based on SDN/NFV, it does not assure the users identity; thus, a challenge for the 5G. An approach for authentication and privacy protection for the 5G small cell vehicular was proposed in \cite{m30}, where the authors specifically presented the idea of a non-authorized assignment design named CLASC that reduces the low-communications overhead. This approach monitors road and vehicle systems and considers the restrictions on authentication and privacy as future tasks in the heterogeneous 5G networks.

In another work, an architecture named blockchain-based trusted authentication (BTA), based on the chain cell was designed for the 5G network by \cite{m31}, which is being applied through the blockchain-based anonymous access (BAA) approach used in  cloud radio over fiber network. Despite the proposed work’s advantages, this study does not discuss the handover challenges among the heterogeneous 5G networks nor their privacy protection aspects.The architecture of the 5G network where the fog computing and radio access network are integrated is devised in\cite{m32}; to achieve Privacy protection, named the F-RAN architecture. Two loosely and highly coupled approaches for computing functions in 5G are devised to address several privacy attacks that identify the attack location among fog nodes in the F-RAN architecture, which does not address the Authentication and challenges and issues in the 5G.

The authentication handover and privacy protection in 5G networks have been discussed in \cite{m5}, where the authors have discussed the application of SDN. In this study, by sharing users’ content among APs, the privacy protection and handover are discussed. Through SDN the 5G heterogeneous cells are managed but sharing the user’s content when a user is passing through heterogeneous networks would lead to the different content transformation of APs, which is followed by overhead and security challenges like data leakage, and a delay in SDN controllers. 
\section{ Conclusion}
The 5G networks with heterogeneous cells and expansion in overlay network coverage are replacing previous generations of mobile networks. A reduction in delay, which is one of the objectives and characteristics of the 5G, is of great importance that can happen with a solid architecture. In this paper, with the assistance of blockchain technology and SDN structure, a new authenticate approach was proposed to protect the privacy of ursers in a faster, safer and more effective manner for the advancement of the 5G network and to provide intelligent control across heterogeneous cells. As results indicated, by removing the repeated manipulations among heterogeneous cells, low latency was obtained for the 5G network. In addtion, with a more lightweight blockchain, instead of applying the POW, the upgraded DPOS consensus algorithm associated with our BC demonstrated to be a better fit for scalability and optimized energy consumption. 
\end{document}